\documentclass[12pt]{article}
\usepackage[english]{babel}
\usepackage{cite}
\usepackage{amsmath,amssymb}
\usepackage{epsfig}

\begin{document}
\thispagestyle{empty}
\hfill DESY 17-219 \\

\vspace*{2.0cm}

\begin{center}
\boldmath
 {\large \bf
  Derivatives of Horn-type hypergeometric functions with respect to their parameters
 }
\unboldmath
\end{center}
 \vspace*{0.8cm}

\begin{center}
{\sc Vladimir~V.~Bytev,$^{a,b,}$\footnote{E-mail: bvv@jinr.ru}
Bernd~A.~Kniehl$^{a,}$\footnote{E-mail: bernd.kniehl@desy.de}
Sven-Olaf Moch$^{a,}$\footnote{E-mail: sven-olaf.moch@desy.de}
} \\
 \vspace*{1.0cm}
{\normalsize $^{a}$ II. Institut f\"ur Theoretische Physik, Universit\"at Hamburg,}\\
{\normalsize Luruper Chaussee 149, 22761 Hamburg, Germany} \\
\bigskip
{\normalsize $^{b}$ Joint Institute for Nuclear Research,} \\
{\normalsize $141980$ Dubna (Moscow Region), Russia}
\end{center}

\begin{abstract}
We consider the derivatives of Horn hypergeometric functions of any number
variables with respect to their parameters.
The derivative of the function in $n$ variables is expressed as a Horn
hypergeometric series of $n+1$ infinite summations depending on the same
variables and with the same region of convergence as for original Horn function.
The derivatives of Appell functions, generalized hypergeometric functions, confluent
and non-confluent Lauricella series and generalized Lauricella series are explicitly presented.
Applications to the calculation of Feynman diagrams are discussed, especially
the series expansion in $\epsilon$ within dimensional regularization.
Connections with other classes of special functions are discussed as well.
\end{abstract}

\newpage

\section{Introduction}
\label{Intro}
In the recent years a lot of attention
\cite{Miller1}--\!\cite{Greynat:2014jsa} has been devoted to hypergeometric series containing the digamma or psi function $\psi(z)$,
\begin{gather}
  \psi(z)=\frac{\Gamma'(z)}{\Gamma(z)}
  \, .
  \label{digamma}
\end{gather}
On the other hand, series with gamma functions are known since long time, see for example the definitions
in the book of Hansen~\cite{hansen}.
Recently, however, some new summations for hypergeometric-type series which
contain digamma functions have been established.
In a series of papers~\cite{Miller1}, \cite{Miller2} and by Cvijovic~\cite{Cvij1}
summation formulae have been derived for hypergeometric-type series which
contain a digamma function as a factor by using certain transformation and
reduction formulae in the theory of Kamp{\'e} de F{\'e}riet
double hypergeometric functions.

Renewed interest in series of hypergeometric-type containing the digamma
function has emerged in connection with derivatives of hypergeometric functions in their parameters.
The first derivatives for some special values of the parameters were already known long time ago~\cite{firstDer1}--\!\cite{firstDer3}.
Later on, Ancarani et al. have found in a series of papers~\cite{Ancarani0}, \cite{Ancarani1} and
\cite{Bujar} the derivatives of Gaussian hypergeometric functions, and
some derivatives of two-variable series, namely the Appell series and four
degenerate confluent series~\cite{Ancarani2}.
Moreover, it has been shown, that the first derivatives of generalized  hypergeometric
functions is expressible in the terms of Kamp{\'e} de F{\'e}riet functions
and, with the same technique, derivatives of the Appell hypergeometric
function was obtained in paper\cite{Sahai}.

In another approach based on the expression of the Pochhammer symbol
\begin{equation}
  (a)_n=\frac{\Gamma(a+n)}{\Gamma(a)}
  \, ,
\label{pochhammer}
\end{equation}
and its reciprocal value the derivative in terms of Stirling numbers was
provided in the papers of Greynat et al.~\cite{Greynat:2013hox,Greynat:2014jsa}.
With such an approach one has the possibility to express some special parameter cases
of Appell or generalized hypergeometric functions in the terms of finite sum
of well-known special function as nested harmonic sums \cite{Moch:2001zr}.
Also, one may formulate the derivatives as series suitable for numerical estimation.

In all cited papers except \cite{Greynat:2013hox,Greynat:2014jsa} the technique of
infinite series resummation~\cite{ShrivastavaResum} is used
\begin{equation}
  \sum_{n=0}^{\infty}\sum_{k=0}^{n}A(k,n)=\sum_{n=0}^{\infty}\sum_{k=0}^{\infty}A(k,n+k)
  \, ,
\label{resum}
\end{equation}
which might not necessarily be correct if applied to series which are
not absolutely convergent.
When the derivative of a hypergeometric function is written as power series in
its arguments of course some discussion on the convergence of the function
obtained is needed.
Only in some special cases explicit formulae for derivatives are presented
(Appell function, some confluent hypergeometric function of two variables),
but there are no explicit results for mixed derivatives of generalized
Kamp{\'e} de F{\'e}riet function.

In various mathematical or physical applications one finds hypergeometric series
which belong to classes of functions different from generalized Kamp{\'e} de
F{\'e}riet functions,
for example, Horn hypergeometric series of two variables, $H_3(a,b,c,x,y)$
(see eq.~(\ref{H3Def}) below for the definition),
where a Pochhammer symbol $(a)_{2m+n}$ with a double summation index is encountered.
This function belongs to the class of generalized Lauricella
series~\cite{ShrivKarls} (see the definition in Sec.~(\ref{LaurGenDef}) below),
and the questions arise how the derivatives of this function look like and
to which class of special functions they belong.

In high-energy physics one has to calculate higher-order Feynman diagrams
for quantum corrections to electroweak and QCD processes.
These are expressible in the form of Mellin-Barnes integrals~\cite{MB1}--\!\cite{MB4},
which depend on the external kinematic invariants, the dimension $D$ of space-time,
and the powers of the propagators.
Upon application of Cauchy's theorem, the Feynman integrals can be converted into linear combinations of Horn-type
hypergeometric series,
\begin{equation}
  \sum_{k_1,\cdots, k_{r+m}=0}^\infty\,
  \prod_{a,b}\,
  \frac{\Gamma(\sum_{i=1}^m {A}_{ai}k_i+{B}_a)}{\Gamma(\sum_{j=1}^r {C}_{bj}k_j+{D}_{b})}\,
  x_1^{k_1} \cdots x_{r+m}^{k_{r+m}}
  \, ,
\label{Horn}
\end{equation}
where $x_i$ are some rational functions of the external kinematic invariants
(e.g., Mandelstam variables) and $A_{ai}, B_a, C_{bj}, D_b$ are linear
functions of the space-time dimension and the propagator powers.
The parameters $A_{ai}, C_{bj}$ do not belong to the set of natural numbers
$\mathbb{N}$ as in the case of generalized Lauricella function,
but could take any integer value.
Within the framework of dimensional regularization, i.e. taking
the space-time parameter $D=4-2\varepsilon$, one has to construct the
so-called $\varepsilon$ expansion of eq.~(\ref{Horn}) over parameter of
dimensional regularization, or just the derivatives of eq.~(\ref{Horn}) with
respect to the  $B_a, D_b$ parameters.
It is very interesting to find explicit formulae for such an derivatives
and the class of functions to which they belong.
These questions provide the motivation for the present paper.

The paper is organized as follows.
We begin in Sec.~\ref{SectN3} where derivatives are considered which affect
one summation index parameter.
As an example, the first order derivative of generalized hypergeometric
functions and the Appell function are presented.
Next, Sec.~\ref{2inexsum} is devoted to the derivative in the case of multiple
summation index parameters and arbitrary derivatives of the well-known
generalized hypergeometric functions and Appell series with respect to their
parameters are discussed.
In Sec.~\ref{sect2n} the derivative in the case of a double summation index
parameter $2n$ is considered, while more the involved cases of multiple
summation index parameters $qn$, $q\in \mathbb{N}$, are given in
Sec.~\ref{qnOneSect} and~\ref{multLaurich}.
As applications, the derivative in the parameters of generalized Lauricella
hypergeometric functions are derived and the derivative of Horn-type
hypergeometric functions in two variables $H_3(a,b,c,x,y)$ is calculated.
Subsequently, Secs.~\ref{SectNeg} and~\ref{SectQNeg} are devoted to the case of an
summation index parameter $qn$, where $q$ is negative and in
Sec.~\ref{SectConv} the region of convergence for such series is discussed.
The main results are collected in Sec.~\ref{mathRes}, where we put the final
equations for derivatives in parameters of hypergeometric functions.
We conclude in Sec.~\ref{conclusion} where we discuss also possible
applications to the calculation of Feynman diagrams.
The Appendix~\ref{hypGenDef} summarizes the definitions of hypergeometric
series used in the paper.

\section{Derivative in parameter with one summation index}
\label{SectN3}
\subsection{Upper parameter derivatives}
\label{sec3}

As a first step we consider the derivative of a hypergeometric function in the
parameter $a$ in the case when the Pochhammer symbol contains only one index
of summation, $(a)_n$. As mentioned above, our calculations in this Section
are similar to the \cite{Miller1}--\!\cite{Bujar}.

The main trick is to consider the derivative of the Pochhammer symbol $(a)_n$.
By using the definition of the digamma function in eq.~(\ref{digamma}), which is
the logarithmic derivative of the gamma function $\Gamma(z)$, and the difference equation
\begin{equation}
\Psi(z+n)-\Psi(z)=\sum_{k=0}^{n-1}\frac{1}{z+k}
\, ,
\label{pochhammerPos}
\end{equation}
we can write the derivative of a Pochhammer symbol in the form
\begin{equation}
\frac{\mathrm{d} (a)_n}{\mathrm{d} a}=(a)_n\biggl[\Psi(a+n)-\Psi(a)\biggr]
=(a)_n\sum_{k=0}^{n-1}\frac{1}{a+k}=(a)_n\frac{1}{a}\sum_{k=0}^{n-1}\frac{(a)_k}{(a+1)_k}
\, .
\label{PochDer}
\end{equation}
For convenience let us write the hypergeometric function in the form
\begin{gather}
F(a)=\sum_{n=0}^\infty B(n) (a)_n \frac{x^n}{n!}
\, .
\label{HypDef}
\end{gather}
Here, we explicitly write the parameter $a$ to be differentiated
and connect it with the summation index $n$ of variable $x$.
The summation over the index $n$ is then explicitly displayed,
but any number of additional summation indices and Pochhammer symbols
are summarized in the coefficient $B(n)$,
\begin{gather}
B(n)=\sum_{m_1\dots m_l}\frac{x_1^{m_1}\dots x_l^{m_l}}{m_1!\dots m_l!}\,
\prod_j\, \frac{(a_j)_{\sum_{i}q_i m_i}}{(b_j)_{\sum_{i}q_i m_i}}
\, .
\end{gather}
We will use this notation and the abbreviation throughout the article.

With the help of eq.~(\ref{PochDer}) and a shift of the summation index $n\to
n+1$  we can write for derivative of function $F(a)$ with respect to the upper
parameter:
\begin{eqnarray}
\frac{\mathrm{d} F(a)}{\mathrm{d} a}=\sum_{n=1}^\infty B(n) (a)_n \frac{x^n}{n!} \frac{1}{a}\sum_{k=0}^{n-1}\frac{(a)_k}{(a+1)_k}
=\sum_{n=0}^\infty B(n+1) (a+1)_n \frac{x^{n+1}}{(n+1)!} \frac{1}{a}\sum_{k=0}^{n}\frac{(a)_k}{(a+1)_k}
\, ,
\end{eqnarray}
and by using the rearrangement formula of summation indices eq.~(\ref{resum})
we obtain the first derivative with respect to an upper one-index parameter:
\begin{eqnarray}
\frac{\mathrm{d} F(a)}{\mathrm{d} a}
=x \sum_{n,k=0}^\infty B(n+k+1)\frac{x^n}{n!}\frac{x^k}{k!}\frac{(1)_k(1)_n}{(2)_{n+k}}\frac{(a+1)_{n+k} (a)_k}{(a+1)_k}
\, .
\label{one-upper}
\end{eqnarray}
For the Gauss hypergeometric function ${}_2F_1(a,b,c,x)$ we obtain \cite{Ancarani0}:
\begin{eqnarray}
\frac{\mathrm{d}}{\mathrm{d} a}\;
{}_{2}F_{1} \left( \begin{array}{c|}
a,b \\
c
\end{array}\, x \right)
&=&\frac{b x }{c}\sum_{k=0}^{\infty}\frac{(a)_k}{(a+1)_k}\sum_{n=0}^\infty\frac{(a+1)_{n+k}(b+1)_{n+k}} {(c+1)_{n+k} }\frac{x^{n+k}}{(n+k+1)!}
\nonumber \\
&=&\frac{b x }{c}\sum_{k=0}^{\infty}\frac{(1)_k (a)_k }{(a+1)_k}\sum_{n=0}^\infty\frac{(1)_n (a+1)_{n+k}(b+1)_{n+k}} {(2)_{n+k} (c+1)_{n+k} }\frac{x^{n} x^k}{n! k!}
\, .
\end{eqnarray}
This hypergeometric series can be understood as a generalized Kamp{\'e} de F{\'e}riet
hypergeometric function (see Sec.~\ref{CamdeFerDef} for definitions):
\begin{gather}
\frac{\mathrm{d}}{\mathrm{d} a}\;
{}_{2}F_{1} \left( \begin{array}{c|}
a,b \\
c
\end{array}\, x \right)
=\frac{b x }{c} F^{2:2;1}_{2:1;0}
\left[ \begin{array}{c}
(a+1,b+1): (1,a);(1) \\
(c+1,2): (a+1);(-)
\end{array}\, x,x \right]
\, .
\end{gather}
It can easily be seen that eq.~(\ref{one-upper}) is suitable for computing the
derivative of any Horn function in one summation index parameter.
For example, one can calculate the derivative of the Appell function $F_1(a,b_1,b_2,c;x,y)$
in the parameter $b_2$, as follows~\cite{Ancarani2}:
\begin{equation}
\frac{\mathrm{d} F_1}{\mathrm{d} b_2}=\frac{a y}{c}\sum_{k=0}^{\infty}\frac{(1)_k (b_2)_k}{(b_2+1)_k}\frac{y^k}{k!}\sum_{n=0}^{\infty}
\frac{(1)_n (b_2+1)_{n+k} }{(2)_{n+k} }\frac{y^n}{n!}
\sum_{m=0}^{\infty}
\frac{ (b_1)_m (a+1) _{m+n+k}}{(c+1)_{m+n+k }}\frac{x^m}{m!}
\, ,
\end{equation}
where the derivative of $F_1$ is now expressed in the terms of the generalized
Lauricella hypergeometric function (see Sec.~\ref{LaurGenDef} for the
definition) as:
\begin{gather}
\frac{\mathrm{d} F_1}{\mathrm{d} b_2}=\frac{ay}{c}
F^{2:1;1;2}_{2:0;0;1}
\left( \begin{tabular}{c}
[a+1:1,1,1; $b_2+1$:0,1,1]:  [$b_1$:1];[1:1];[1:1;$b_2$;1]
\\
\, [c+1:1,1,1;  2:0,1,1]: [-];[-];[$b_2$+1,1]
\end{tabular}\, x,y,y \right)
\, .
\end{gather}

The corresponding expression for the derivative of $F_1$ in the parameter $b_1$ can then be obtained via exchange rule:
\begin{equation}
\frac{\mathrm{d} F_1(a,b_1,b_2,c;x,y)}{\mathrm{d} b_1}=\frac{\mathrm{d} F_1(a,b_1,b_2,c;x,y)}{\mathrm{d} b_2}\biggr|_{b_1\leftrightarrow b_2,x\leftrightarrow y}
\, .
\end{equation}
With similar manipulations one can compute the derivatives in an upper
one-index parameter for the last three Lauricella functions:
\begin{equation}
\frac{\mathrm{d} F_2}{\mathrm{d} b_2}=
\frac{a y}{c_2}\sum_{k=0}^{\infty}\frac{(1)_k (b_2)_k}{(b_2+1)_k}\frac{y^k}{k!}\sum_{n=0}^{\infty}
\frac{(1)_n (b_2+1)_{n+k} }{(2)_{n+k} (c_2+1)_{n+k}}\frac{y^n}{n!}
\sum_{m=0}^{\infty}
\frac{ (b_1)_m (a+1) _{m+n+k}}{(c_1)_{m }}\frac{x^m}{m!}
\, ,
\end{equation}
\begin{equation}
\frac{\mathrm{d} F_3}{\mathrm{d} b_2}=
\frac{a_2 y}{c}\sum_{k=0}^{\infty}\frac{(1)_k (b_2)_k}{(b_2+1)_k}\frac{y^k}{k!}\sum_{n=0}^{\infty}
\frac{(1)_n (b_2+1)_{n+k}  (a_2+1)_{n+k}}{(2)_{n+k} }\frac{y^n}{n!}
\sum_{m=0}^{\infty}
\frac{ (b_1)_m (a_1) _{m}}{(c+1)_{m+n+k }}\frac{x^m}{m!}
\, .
\end{equation}
By using the symmetries of hypergeometric functions, we can then find the
other derivatives with respect to upper one-index parameters:
\begin{eqnarray}
\frac{\mathrm{d} F_2(a,b_1,b_2,c_1,c_2;x,y)}{\mathrm{d} b_1}&=&\frac{\mathrm{d} F_2(a,b_1,b_2,c_1,c_2;x,y)}{\mathrm{d} b_2}\biggr|_{b_1\leftrightarrow b_2,
c_1\leftrightarrow c_2,x\leftrightarrow y}
\, ,
\nonumber\\
\frac{\mathrm{d} F_3(a_1,a_2,b_1,b_2,c;x,y)}{\mathrm{d} b_1}&=&\frac{\mathrm{d} F_3(a_1,a_2,b_1,b_2,c;x,y)}{\mathrm{d} b_2}\biggr|_{a_1\leftrightarrow a_2,
b_1\leftrightarrow b_2,
x\leftrightarrow y}
\, ,
\nonumber\\
\frac{\mathrm{d} F_3(a_1,a_2,b_1,b_2,c;x,y)}{\mathrm{d} a_2}&=&\frac{\mathrm{d} F_3(a_1,a_2,b_1,b_2,c;x,y)}{\mathrm{d} b_2}\biggr|_{a_1\leftrightarrow b_1,
a_2\leftrightarrow b_2}
\, ,
\nonumber\\
\frac{\mathrm{d} F_3(a_1,a_2,b_1,b_2,c;x,y)}{\mathrm{d} a_1}&=&\frac{\mathrm{d} F_3(a_1,a_2,b_1,b_2,c;x,y)}{\mathrm{d} b_2}\biggr|_{a_1\leftrightarrow b_2,
b_1\leftrightarrow a_2,
x\leftrightarrow y}
\, .
\end{eqnarray}

\subsection{Lower parameter derivative}

If the derivative is acting on a lower parameter eq.~(\ref{PochDer}) changes
to the derivative of a reciprocal Pochhammer symbol:
\begin{equation}
\frac{\mathrm{d}}{\mathrm{d} b}\frac{1}{ (b)_n}=\frac{1}{(b)_n}\biggl[\Psi(b)-\Psi(b+n)\biggr]
=-\frac{1}{(b)_n}\sum_{k=0}^{n-1}\frac{1}{b+k}=-\frac{1}{(b)_n}\frac{1}{b}\sum_{k=0}^{n-1}\frac{(b)_k}{(b+1)_k}
\, .
\end{equation}
Then, using a short-hand notation for the hypergeometric function similar to
eq.~(\ref{HypDef}), i.e.
\begin{gather}
{F}(b)=\sum_{n=0}^\infty B(n) \frac{1}{(b)_n} \frac{x^n}{n!}
\, ,
\end{gather}
one can express derivatives with respect to the lower one-index parameter as:
\begin{eqnarray}
\frac{\mathrm{d} {F}(b)}{\mathrm{d} b}=-\sum_{n=1}^\infty D(n)\frac{x^n}{n!} \frac{1}{(b)_n} \frac{1}{b}\sum_{k=0}^{n-1}\frac{(b)_k}{(b+1)_k}
=-\sum_{n=0}^\infty D(n+1)\frac{x^{n+1}}{(n+1)!} \frac{1}{(b)_{n+1}} \frac{1}{b}\sum_{k=0}^{n}\frac{(b)_k}{(b+1)_k}
\, ,
\nonumber
\end{eqnarray}
and after rearrangement of the sums one obtains the following formula
\begin{eqnarray}
  \frac{\mathrm{d} {F}(b)}{\mathrm{d} b}
  =-\frac{x}{b^2}\sum_{n,k=0}^\infty{D(n+k+1)}\frac{x^n x^k}{n! k!} \frac{(1)_n (1)_k}{(2)_{n+k}} \frac{1}{  {(b+1)_{n+k}} }\frac{(b)_k}{(b+1)_k}
  \, .
  \label{one-lower}
 \end{eqnarray}
As an example, we obtain for the Gauss hypergeometric function
${}_2F_1(a,b,c,x)$ the result~\cite{Ancarani0},
\begin{eqnarray}
\frac{\mathrm{d}}{\mathrm{d} c}\;
{}_{2}F_{1} \left( \begin{array}{c|}
a,b \\
c
\end{array}\, x \right)
&=&-\frac{ a b x }{c^2}\sum_{k=0}^{\infty}\frac{(c)_k}{(c+1)_k}\sum_{n=0}^\infty\frac{(a+1)_{n+k}(b+1)_{n+k}} {(c+1)_{n+k} }\frac{x^{n+k}}{(n+k+1)!}
\nonumber \\
&=&-\frac{a b x }{c^2}\sum_{k=0}^{\infty}\frac{(1)_k (c)_k }{(c+1)_k}\sum_{n=0}^\infty\frac{(1)_n (a+1)_{n+k}(b+1)_{n+k}} {(2)_{n+k} (c+1)_{n+k} }\frac{x^{n} x^k}{n! k!}
\, ,
\end{eqnarray}
which is a series that can be expressed in terms of the generalized  Kamp{\'e} de F{\'e}riet-type function:
\begin{gather}
\frac{\mathrm{d}}{\mathrm{d} c}\;
{}_{2}F_{1} \left( \begin{array}{c|}
a,b \\
c
\end{array}\, x \right)
=-\frac{a b x }{c^2}
 F^{2:1;1}_{2:1;1}
\left[ \begin{array}{c}
(a+1,b+1): (1,c);(1) \\
(c+1,2): (c+1);(-)
\end{array}\, x,x \right]
\, .
\end{gather}

Again, one can use eq.~(\ref{one-lower}) also for differentiating any
hypergeometric function in a lower parameter with dependence on one summation
index only.
For the derivatives of the functions $F_2$ and $F_4$ in the lower parameters  $c_1$, $c_2$ we find:
\begin{equation}
\frac{\mathrm{d} F_2}{\mathrm{d} c_2}=-\frac{y a b_2}{c_2^2}\sum_{k=0}^{\infty}\frac{(1)_k (c_2)_k}{(c_2+1)_k}\frac{y^k}{k!}\sum_{n=0}^{\infty}
\frac{(1)_n (b_2+1)_{n+k}  }{(2)_{n+k} (c_2+1)_{n+k}}\frac{y^n}{n!}
\sum_{m=0}^{\infty}
\frac{ (b_1)_m (a+1)_{m+n+k}}{(c_1)_{m }}\frac{x^m}{m!}
\, ,
\nonumber
\end{equation}
\begin{equation}
\frac{\mathrm{d} F_4}{\mathrm{d} c_2}=-\frac{y a b}{c_2^2}\sum_{k=0}^{\infty}\frac{(1)_k (c_2)_k}{(c_2+1)_k}\frac{y^k}{k!}\sum_{n=0}^{\infty}
\frac{(1)_n   }{(2)_{n+k} (c_2+1)_{n+k}}\frac{y^n}{n!}
\sum_{m=0}^{\infty}
\frac{ (a+1)_{m+n+k} (b+1)_{m+n+k}}{(c_1)_{m }}\frac{x^m}{m!}
\, ,
\nonumber
\end{equation}
\begin{eqnarray}
\frac{\mathrm{d} F_2(a,b_1,b_2,c_1,c_2;x,y)}{\mathrm{d} c_1}&=&\frac{\mathrm{d} F_2(a,b_1,b_2,c_1,c_2;x,y)}{\mathrm{d} c_2}\biggr|_{b_1\leftrightarrow b_2,
c_1\leftrightarrow c_2,x\leftrightarrow y}
\, ,
\nonumber\\
\frac{\mathrm{d} F_4(a,b,c_1,c_2,c;x,y)}{\mathrm{d} c_1}&=&\frac{\mathrm{d} F_4(a,b,c_1,c_2,c;x,y)}{\mathrm{d} c_2}\biggr|_{c_1\leftrightarrow c_2,x\leftrightarrow y}
\, .
\end{eqnarray}
Also these derivatives are then expressible in the terms of generalized Lauricella hypergeometric functions:
\begin{gather}
\frac{\mathrm{d} F_2}{\mathrm{d} c_2}=-\frac{y a b_2}{c_2^2}
F^{2:1;1;2}_{2:1;0;1}
\left( \begin{tabular}{c}
[a+1:1,1,1; $b_2$+1:0,1,1]:  [$b_1$:1];[1:1];[1:1;$c_2$:1]
\\
\, [$c_2$+1:0,1,1;  2:0,1,1]: [$c_1$:1];[-];[$c_2$+1:1]
\end{tabular}\, x,y,y \right)
\, ,
\end{gather}
\begin{gather}
\frac{\mathrm{d} F_4}{\mathrm{d} c_2}=-\frac{y a b}{c_2^2}
F^{2:0;1;2}_{2:1;0;1}
\left( \begin{tabular}{c}
[a+1:1,1,1; $b$+1:1,1,1]:  [-];[1:1];[1:1;$c_2$:1]
\\
\, [$c_2$+1:0,1,1;  2:0,1,1]: [c:1];[-];[$c_2$+1,1]
\end{tabular}\, x,y,y \right)
\, .
\end{gather}
From eqs.~(\ref{one-upper}), (\ref{one-lower}) and the definition of generalized Lauricella series in Sec.~\ref{LaurGenDef}
one can see that only the first derivatives of generalized hypergeometric functions in a lower parameter
can be written in terms of generalized Kamp{\'e} de F{\'e}riet functions in the two variables.
On the other hand, mixed or higher order derivatives of generalized hypergeometric functions, as
well as Appell hypergeometric series can only be stated in terms of generalized Lauricella series.

\section{Derivative in parameter with multiple summation indices}
\label{2inexsum}
\subsection{Upper parameter derivative}
\label{multSummsect}
Next, we consider the derivative in the case of multiple summation indices in Pochhammer symbol $(a)_{n_1+n_2+....}$.
Such sums arise in the calculation of mixed derivatives of
$pFq$ or the first derivative of an Appell function in two summation index parameters
$(a)_{n_1+n_2}$.
In that case we factorize the Pochhammer symbol as a product of two terms with
one summation index each,
\begin{gather}
(a)_{n_1+n_2}=(a+n_1)_{n_2}(a)_{n_1}
\, ,
\label{splitting}
\end{gather}
or in the case of multiple summation indices,
\begin{gather}
(a)_{\sum_{\lambda=1}^{\phi} n_\lambda}=(a+\sum_{\lambda=1}^{\phi-1}n_\lambda)_{n_\phi}
(a+\sum_{\lambda=1}^{\phi-2}n_\lambda)_{n_{\phi-1}}\dots (a)_{n_1}=\prod_{r=1}^\phi(a+\sum_{\lambda=1}^{r-1}n_\lambda)_{n_r}
\, .
\label{splittingmultiple}
\end{gather}
Upon expressing the hypergeometric function in the form
\begin{gather}
F(a)=\sum_{m,n=0}^\infty B(n,m) (a)_{m+n} \frac{x^n y^m}{n! m!}
\, ,
\end{gather}
and applying eq.~(\ref{splitting}) to factorize the Pochhammer symbol
together with the results for the derivatives in one-index parameters
from Sec.~\ref{sec3} one obtains:
\begin{eqnarray}
\frac{\mathrm{d} F(a)}{\mathrm{d} a}&=&y\sum_{k,n,m=0}^\infty B(n,m+k+1)\frac{(1)_k (1)_m}{(2)_{m+k}} \frac{(a)_k  (a+1)_{m+n+k}}{(a+1)_{k}}\frac{x^n y^m y^k}{n! m! k!}
\nonumber \\
&+&x\sum_{k,n,m=0}^\infty B(n+k+1,m)\frac{(1)_k (1)_n}{(2)_{n+k}} \frac{(a)_{m+k}  (a+1)_{m+n+k}}{(a+1)_{m+k}}\frac{x^n y^m x^k}{n! m! k!}
\, .
\end{eqnarray}
As an example, the derivative of the Appell hypergeometric function $F_1$ in
the parameter $a$ with two summation indices $m+n$ reads~\cite{Ancarani2},
\begin{eqnarray}
\frac{\mathrm{d} F_1}{\mathrm{d} a}&=&\frac{y b_1}{c}\sum_{k,n,m=0}^\infty\frac{(b_1+1)_{m+k} (b_2)_n}{(c+1)_{m+n+k}}\frac{(1)_k (1)_m}{(2)_{m+k}} \frac{(a)_k  (a+1)_{m+n+k}}{(a+1)_{k}}\frac{x^n y^m y^k}{n! m! k!}
\nonumber \\
&+&\frac{x b_2}{c}\sum_{k,n,m=0}^\infty\frac{(b_1)_{m} (b_2+1)_{n+k}}{(c+1)_{m+n+k}}\frac{(1)_k (1)_n}{(2)_{n+k}} \frac{(a)_{m+k}  (a+1)_{m+n+k}}{(a+1)_{m+k}}\frac{x^n y^m x^k}{n! m! k!}
\, ,
\end{eqnarray}
or, alternatively, in terms of the generalized Lauricella series,
\begin{eqnarray}
\frac{\mathrm{d} F_1}{\mathrm{d} a}&=&\frac{y b_1}{c}
F^{2:1;1;2}_{2:0;0;1}
\left( \begin{tabular}{c}
[a+1:1,1,1; $b_1$+1:1,0,1]:  [1:1];[$b_2$:1];[1:1;$a$:1]
\\
\, [$c$+1:1,1,1;  2:1,0,1]: [-];[-];[a+1:1]
\end{tabular}\, y,x,y \right)
\nonumber \\
&+&\frac{x b_2}{c}
F^{3:1;1;1}_{3:0;0;0}
\left( \begin{tabular}{c}
[a+1:1,1,1; $a$:1,0,1; $b_2$+1:0,1,1]:  [$b_1$:1];[1:1];[1:1]
\\
\, [$c$+1:1,1,1;  2:0,1,1; a+1:1,0,1]: [-];[-];[-]
\end{tabular}\, y,x,x \right)
\, .
\end{eqnarray}
In the case of hypergeometric function with the multiple summation indices
\begin{gather}
F(a)=\sum_{n_1,\dots,n_\phi=0}^\infty B(n_1,\dots,n_\phi) (a)_{\sum_{r=1}^\phi n_\lambda}\prod_{r=1}^\phi \frac{x^{n_r} }{n_r!}
\, ,
\end{gather}
using eq.~(\ref{splittingmultiple}) and the parametric derivative with multiple summation indices
\begin{gather}
\frac{\mathrm{d}}{\mathrm{d} a}(a+\sum_{\lambda=1}^{\xi-1}n_\lambda)_{n_\xi}=(a+\sum_{\lambda=1}^{\xi-1}n_\lambda)_{n_\xi}
\sum_{k=0}^{n_\xi-1}\frac{1}{a+\sum_{\lambda=1}^{\xi-1}n_\lambda+k},
\end{gather}
one obtains for case with multiple summation indices
\begin{eqnarray}
\frac{\mathrm{d} F(a)}{\mathrm{d} a}&=&\sum_{k,n_1,\dots,n_\phi=0}^\infty \sum_{\xi=1}^\phi x_\xi B(n_1,\dots,n_\xi+k+1,\dots,n_\phi)\frac{(1)_k (1)_{n_\xi}}{(2)_{n_\xi+k}}
\prod_{r=1}^\phi\frac{x_r^{n_r} x_\xi^{k}}{n_r! k!}
\nonumber
\\
&\times&
\prod_{r=1}^{\xi-1}\frac{(a)_{\sum_{\lambda=1}^{r}n_\lambda}  }{ (a)_{\sum_{\lambda=1}^{r-1}n_\lambda} }
\prod_{r=\xi+1}^{\phi}\frac{(a+1)_{\sum_{\lambda=1}^{r}n_\lambda+k}  }{ (a+1)_{\sum_{\lambda=1}^{r-1}n_\lambda+k} }
\nonumber\\
&\times&
\frac{(a+1)_{\sum_{\lambda=1}^{\xi}n_\lambda+k} }{(a+1)_{\sum_{\lambda=1}^{\xi-1}n_\lambda+k} }
\frac{(a)_{\sum_{\lambda=1}^{\xi-1}n_\lambda+k} }{(a)_{\sum_{\lambda=1}^{\xi-1}n_\lambda} }
\, .
\label{multSumEq}
\end{eqnarray}

\subsection{Lower parameter derivative}
The same trick can be applied to the derivative acting on a lower parameter with multiple summation indices, i.e., the case
\begin{gather}
F(b)=\sum_{m,n=0}^\infty B(n,m) \frac{1}{(b)_{m+n}} \frac{x^n y^m}{n! m!}
\, .
\end{gather}
Upon factorizing the reciprocal Pochhammer symbol similar to
eq.~(\ref{splitting}) as
\begin{gather}
\frac{1}{(b)_{m+n}}=\frac{1}{(b+m)_{n}}\frac{1}{(b)_{m}}
\, ,
\end{gather}
or in the case of multiple summation indices as,
\begin{gather}
\frac{1}{(b)_{\sum_{\lambda=1}^{\phi} n_\lambda}}=\prod_{r=1}^\phi\frac{1}{(b+\sum_{\lambda=1}^{r-1}n_\lambda)_{n_r}}
\, ,
\label{lowerMultSumm}
\end{gather}
one can express the derivative in a lower double summation index parameter of a hypergeometric function in terms of hypergeometric series:
\begin{eqnarray}
\frac{\mathrm{d} F(b)}{\mathrm{d} b}&=&-y\sum_{k,n,m=0}^\infty B(n,m+k+1)\frac{(1)_k (1)_m}{(2)_{m+k}} \frac{1}{b^2}\frac{(b)_k  }{(b+1)_{m+n+k} (b+1)_{k}}\frac{x^n y^m y^k}{n! m! k!}
\nonumber \\
&-&x\sum_{k,n,m=0}^\infty B(n+k+1,m)\frac{(1)_k (1)_n}{(2)_{n+k}} \frac{1}{b^2} \frac{(b)_{m+k}  }{(b+1)_{m+n+k} (b+1)_{m+k}}\frac{x^n y^m x^k}{n! m! k!}
\, .
\end{eqnarray}
As an example for illustration we present the derivative of the
Appell function $F_3$ in its lower parameter $c$ with summation indices $m+n$,
see~\cite{Ancarani2},
\begin{eqnarray}
\frac{\mathrm{d} F_3}{\mathrm{d} c}&=&-\frac{y a_1 b_1}{c^2}\sum_{k,n,m=0}^\infty\frac{(a_1+1)_{m+k} (a_2)_n (b_1+1)_{m+k} (b_2)_n (c)_k}{(c+1)_{m+n+k}(c+1)_k}\frac{(1)_k (1)_m}{(2)_{m+k}} \frac{x^n y^m y^k}{n! m! k!}
\nonumber \\
&-&\frac{x a_2 b_2}{c^2}\sum_{k,n,m=0}^\infty\frac{(a_1)_{m} (a_2+1)_{n+k} (b_1)_m (b_2+1)_{n+k}(c)_{m+k}}{(c+1)_{m+n+k}(c+1)_{m+k}}\frac{(1)_k (1)_n}{(2)_{n+k}} \frac{x^n y^m x^k}{n! m! k!}
\, ,
\end{eqnarray}
or, alternatively, in terms of the generalized Lauricella series:
\begin{eqnarray}
\frac{\mathrm{d} F_3}{\mathrm{d} c}&=&-\frac{y a_1b_1}{c^2}
F^{2:1;2;2}_{2:0;0;1}
\left( \begin{tabular}{c}
[$a_1$+1:1,0,1; $b_1$+1:1,0,1]:  [1:1];[$a_2$:1;$b_2$:1];[1:1;$c$:1]
\\
\, [$c$+1:1,1,1;  2:1,0,1]: [-];[-];[c+1:1]
\end{tabular}\, y,x,y \right)
\nonumber \\
&-&\frac{x a_2 b_2}{c^2}
F^{3:2;1;1}_{3:0;0;0}
\left( \begin{tabular}{c}
[$a_2$+1:0,1,1; $b_2$+1:0,1,1; $c$:1,0,1]:  [$a_1$:1;$b_1$:1];[1:1];[1:1]
\\
\, [$c$+1:1,1,1;  2:0,1,1; c+1:1,0,1]: [-];[-];[-]
\nonumber
\end{tabular}\, y,x,x \right)
\, .
\end{eqnarray}

In complete analogy to the case of the previous subsection for hypergeometric
series with a lower parameter containing multiple summation indices
\begin{gather}
F(b)=\sum_{n_1,\dots,n_\phi=0}^\infty B(n_1,\dots,n_\phi) \frac{1}{(b)_{\sum_{r=1}^\phi n_\lambda}}\prod_{r=1}^\phi \frac{x^{n_r} }{n_r!}
\, ,
\end{gather}
using eq.~(\ref{lowerMultSumm}) and the derivative in a lower parameter with multiple summation indices
\begin{gather}
\frac{\mathrm{d}}{\mathrm{d} b}\frac{1}{(b+\sum_{\lambda=1}^{\xi-1}n_\lambda)_{n_\xi}}=-\frac{1}{(b+\sum_{\lambda=1}^{\xi-1}n_\lambda)_{n_\xi}}
\frac{1}{b+\sum_{\lambda=1}^{\xi-1}n_\lambda}
\sum_{k=0}^{n_\xi-1}\frac{(b+\sum_{\lambda=1}^{\xi-1}n_\lambda)_k}{(b+\sum_{\lambda=1}^{\xi-1}n_\lambda+1)_k}
\, ,
\end{gather}
one obtains for derivative in a lower parameter,
\begin{eqnarray}
\frac{\mathrm{d} F(b)}{\mathrm{d} b}&=&-\frac{1}{b^2}\sum_{k,n_1,\dots,n_\phi=0}^\infty \sum_{\xi=1}^\phi x_\xi B(n_1,\dots,n_\xi+k+1,\dots,n_\phi)\frac{(1)_k (1)_{n_\xi}}{(2)_{n_\xi+k}}
\prod_{r=1}^\phi\frac{x_r^{n_r} x_\xi^{k}}{n_r! k!}
\nonumber
\\
&\times&
\prod_{r=1}^{\xi-1}\frac{(b)_{\sum_{\lambda=1}^{r-1}n_\lambda}  }{ (b)_{\sum_{\lambda=1}^{r}n_\lambda} }
\prod_{r=\xi+1}^{\phi}\frac{(b+1)_{\sum_{\lambda=1}^{r-1}n_\lambda+k}  }{ (b+1)_{\sum_{\lambda=1}^{r}n_\lambda+k} }
\nonumber\\
&\times&
\frac{(b)_{\sum_{\lambda=1}^{\xi-1}n_\lambda} }{(b+1)_{\sum_{\lambda=1}^{\xi}n_\lambda+k} }
\frac{(b)_{\sum_{\lambda=1}^{\xi-1}n_\lambda+k} }{(b+1)_{\sum_{\lambda=1}^{\xi-1}n_\lambda+k} }
\, .
\label{multSumEqLow}
\end{eqnarray}

With the relations presented thus far,
well-known hypergeometric functions such as the generalized hypergeometric,
Appell or Lauricella series can be written in terms of generalized
Lauricella series with the summation coefficients
$\theta_1^{{(1)}},\psi_1^{{(1)}},\phi_1^{{(1)}},\delta_1^{{(1)}}\dots
\theta_A^{{(n)}},\psi_C^{{(n)}},\phi_{B^{(n)}}^{{(n)}},\delta_{D^{(n)}}^{{(n)}}$,
(see Sec.~\ref{LaurGenDef} for definitions) taking the values $1$ or $0$.
Moreover, from eqs.~(\ref{multSumEq}) and (\ref{multSumEqLow}) one can see
that any derivative of these functions in one of their parameters can also be expressed in terms
of generalized Lauricella series with the values of the summation coefficients being $1$ or $0$.
As a more general statement we note that if the initial function can be
expressed as a generalized Lauricella series with summation coefficients from
the alphabet $\{0,1\}$, then any $n$-th derivative of this initial function
can be expressed within the class of the same functions.
The number of variables of $n$-th derivative of an initial function with $m$ variable is $n+m$.

\section{Derivative in parameter with summation index  $2n$}
\label{sect2n}
\subsection{Upper parameter derivative}
\label{up2n}
In some hypergeometric series one encounters the occurrence of summation
indices with a factor two, i.e., $(a)_{2n}$, such that
\begin{gather}
  F(a)=\sum_{n=0}^\infty B(m,n) (a)_{2n} \frac{x^n}{n!}
  \, .
\end{gather}
The simplest realizations are the case of Horn hypergeometric functions in two
variables (see Sec.~\ref{HornDef} for some examples).
In this case one can use eq.~(\ref{PochDer}) for the derivative of the Pochhammer symbol
\begin{equation}
  \frac{\mathrm{d} (a)_{2n}}{\mathrm{d} a}
  =(a)_{2n}\sum_{k=0}^{2n-1}\frac{1}{a+k}=(a)_{2n}\frac{1}{a}\sum_{k=0}^{2n-1}\frac{(a)_k}{(a+1)_k}
  \, ,
\label{PochDer2n}
\end{equation}
together with a rearrangement of the summation formula in eq.~(\ref{resum})
which splits in two terms due to the upper summation limit at $2n+1$,
\begin{equation}
  \sum_{n=0}^{\infty}\sum_{k=0}^{2n+1}A(k,n)=\sum_{n=0}^{\infty}\sum_{k=0}^{\infty}(A(2k,n+k)+A(2k+1,n+k))
  \, .
\label{resum2n}
\end{equation}
For the hypergeometric function with derivative in a parameter $a$ with a summation index ${2n}$,
\begin{gather}
  F(a)=\sum_{m,n=0}^\infty B(n) (a)_{2n} \frac{x^n}{n!}
  \, ,
\end{gather}
one obtains
\begin{eqnarray}
\frac{\mathrm{d} F(a)}{\mathrm{d} a}&=&x\sum_{k,n=0}^\infty B(n+k+1)\frac{(1)_k (1)_n}{(2)_{n+k}}(a+1){(a+2)_{2n+2k}}
\nonumber\\
&\times&\biggl(\frac{a}{a+1} \frac{(a+1)_{2n}  }{(a+2)_{2n}}+\frac{(a)_{2n}  }{(a+1)_{2n}}\biggr)\frac{x^n x^k}{n! k!}
\, .
\end{eqnarray}

\subsection{Lower parameter derivative}
With the same procedure as in Sec.~\ref{up2n} for the derivative in a lower
parameter of the type $(b)_{2n}$ one obtains
\begin{eqnarray}
\frac{\mathrm{d} F(b)}{\mathrm{d} b}&=&-x\sum_{k,n=0}^\infty B(n+k+1)\frac{(1)_k (1)_n}{(2)_{n+k}}\frac{1}{b^2(b+1)(b+2)_{2n+2k}}
\nonumber\\
&\times&\biggl(\frac{b}{b+1} \frac{(b+1)_{2k}  }{(b+2)_{2k}}+\frac{(b)_{2k}  }{(b+1)_{2k}}\biggr)\frac{x^n x^k}{n! k!}
\, .
\end{eqnarray}

\section{Derivative in parameter with summation index $qn$, $q\in \mathbb{N}$}
\label{qnOneSect}
\subsection{Upper parameter derivative}
Here we consider the case when the summation index has a positive integer
coefficient $qn$ with $q\in \mathbb{N}$,
\begin{gather}
  F(a)=\sum_{n=0}^\infty B(n) (a)_{qn} \frac{x^n}{n!}
  \, .
\end{gather}
The particular case of $q=2$ has been dealt with previously in Sec.~\ref{sect2n}.
The derivative of the Pochhammer symbol in this case reads
\begin{equation}
  \frac{\mathrm{d} (a)_{qn}}{\mathrm{d} a}
  =(a)_{qn}\sum_{k=0}^{qn-1}\frac{1}{a+k}=(a)_{qn}\frac{1}{a}\sum_{k=0}^{qn-1}\frac{(a)_k}{(a+1)_k}
  \, ,
\label{PochDerqn}
\end{equation}
and with the help of an extension of the resummation formula in eq.~(\ref{resum2n}) to the case of positive integer factors $q$,
\begin{equation}
  \sum_{n=0}^{\infty}\sum_{k=0}^{qn+q-1}A(k,n)=\sum_{n=0}^{\infty}\sum_{k=0}^{\infty}\sum_{\lambda=0}^{q-1}A(qk+\lambda,n+k)
  \, ,
\label{resumqn}
\end{equation}
one obtains the derivative in an upper parameter with a summation index of
type $qn$ as
\begin{eqnarray}
  \frac{\mathrm{d} F(a)}{\mathrm{d} a}&=&x\sum_{\lambda=0}^{q-1}\frac{1}{a+\lambda}\frac{\Gamma(a+q)}{\Gamma(a)}\sum_{k,n=0}^\infty B(n+k+1)\frac{(1)_k (1)_n}{(2)_{n+k}}
  \nonumber\\
  &\times&
  \frac{{(a+q)_{qn+qk}}(a+\lambda)_{qk}  }{(a+\lambda+1)_{qk}}\frac{x^n x^k}{n! k!}
  \, .
 \label{qnsummEqOne}
\end{eqnarray}

\subsection{Lower parameter derivative}
For the same summation index $qn$ but appearing in a lower parameter we obtain
for derivative in this parameter
\begin{eqnarray}
\frac{\mathrm{d} F(b)}{\mathrm{d} b}&=&-x\sum_{\lambda=0}^{q-1}\frac{1}{b+\lambda}\frac{\Gamma(b)}{\Gamma(b+q)}\sum_{k,n=0}^\infty B(n+k+1)\frac{(1)_k (1)_n}{(2)_{n+k}}
\nonumber\\
&\times&
 \frac{(b+\lambda)_{qk}  }{{(b+q)_{qn+qk}}(b+\lambda+1)_{qk}}\frac{x^n x^k}{n! k!}
 \, .
\end{eqnarray}

\section{Derivative in parameter with multiple summation indices $q_\lambda n_\lambda$, $q_\lambda \in \mathbb{N}$}
\label{multLaurich}
\subsection{Upper parameter derivative}
\label{multQnSect}
By exploiting the previous results, in particular eqs.~(\ref{multSumEq}) and (\ref{qnsummEqOne})
one can obtain the derivative of hypergeometric functions in a parameter with
multiple summation indices $q_\lambda n_\lambda$, where $q_\lambda \in \mathbb{N}$,
\begin{gather}
  F(a)=\sum_{n_1,\dots,n_\phi=0}^\infty B(n_1,\dots,n_\phi) (a)_{\sum_{r=1}^\phi q_\lambda n_\lambda}\prod_{r=1}^\phi \frac{x^{n_r} }{n_r!}
  \, ,
\end{gather}
in the form
\begin{eqnarray}
\frac{\mathrm{d} F(a)}{\mathrm{d} a}&=&\sum_{k,n_1,\dots,n_\phi=0}^\infty \sum_{\xi=1}^\phi \sum_{\gamma=0}^{q_\xi-1} x_\xi B(n_1,\dots,n_\xi+k+1,\dots,n_\phi)\frac{(1)_k (1)_{n_\xi}}{(2)_{n_\xi+k}}
\prod_{r=1}^\phi\frac{x_r^{n_r} x_\xi^{k}}{n_r! k!}
\nonumber
\\
&\times&
\frac{\Gamma(a+q_\xi)}{\Gamma(a)(a+\gamma)}
\prod_{r=1}^{\xi-1}\frac{(a)_{\sum_{\lambda=1}^{r} q_\lambda n_\lambda}  }{ (a)_{\sum_{\lambda=1}^{r-1} q_\lambda n_\lambda} }
\prod_{r=\xi+1}^{\phi}\frac{(a+q_\xi)_{\sum_{\lambda=1}^{r} q_\lambda n_\lambda+q_\xi k}  }{ (a+q_\xi)_{\sum_{\lambda=1}^{r-1} q_\lambda n_\lambda+q_\xi k} }
\nonumber\\
&\times&
\frac{(a+q_\xi)_{\sum_{\lambda=1}^{\xi}q_\lambda n_\lambda+q_\xi k} }{(a)_{\sum_{\lambda=1}^{\xi-1}q_\lambda n_\lambda} }
\frac{(a+\gamma)_{\sum_{\lambda=1}^{\xi-1}q_\lambda n_\lambda+q_\xi k} }{(a+\gamma+1)_{\sum_{\lambda=1}^{\xi-1}q_\lambda n_\lambda+q_\xi k} }
\, .
\label{multSumEqUp}
\end{eqnarray}
As an example, we calculate the derivative of the Horn function
$H_3(a,b,c,x,y)$ (see Sec.~\ref{HornDef} for definitions) in its upper
parameter $a$ with summation indices $2m+n$. This proceeds as
\begin{eqnarray}
\frac{\mathrm{d} H_3(a)}{\mathrm{d} a}&=&\frac{y b}{c}\sum_{m,n,k=0}^\infty \frac{(1)_k (1)_{n}}{(2)_{n+k}}\frac{x^m y^n y^k}{m!n!k!}
\frac{(a)_{2m+k}(1+a)_{2m+n+k}(b+1)_{n+k}}{(1+a)_{2m+k} (c+1)_{m+n+k}}
\nonumber
\\
&+&\frac{x(1+a) }{c}\sum_{m,n,k=0}^\infty \frac{(1)_k (1)_{m}}{(2)_{m+k}}\frac{x^m y^n x^k}{m!n!k!}
\frac{(a)_{2k}(2+a)_{2m+n+2k}(b)_{n}}{(1+a)_{2k} (c+1)_{m+n+k}}
\nonumber
\\
&+&\frac{x a}{c}\sum_{m,n,k=0}^\infty \frac{(1)_k (1)_{m}}{(2)_{m+k}}\frac{x^m y^n x^k}{m!n!k!}
\frac{(1+a)_{2k}(2+a)_{2m+n+2k}(b)_{n}}{(2+a)_{2k} (c+1)_{m+n+k}}
\, ,
\end{eqnarray}
which can be written as a sum of three generalized Lauricella series:
\begin{eqnarray}
\frac{\mathrm{d} H_3(a)}{\mathrm{d} a}
&=&\frac{yb}{c}
F^{3:0;1;1}_{3:0;0;0}
\left( \begin{tabular}{c}
[a+1:2,1,1; b+1:0,1,1;a:2,0,1]: [-];[1:1];[1:1]
\\
\, [c+1:1,1,1; a+1:2,0,1; 2:0,1,1]: [-];[-];[-]
\end{tabular}\, x,y,y \right)
\nonumber
\\
&+&\frac{x(1+a)}{c}
F^{1:1;1;2}_{2:0;0;1}
\left( \begin{tabular}{c}
[a+2:2,1,2]: [1:1];[b:1];[1:1;a:2]
\\
\, [c+1:1,1,1;  2:1,0,1]: [-]; [-];[a+1:2]
\end{tabular}\, x,y,x \right)
\nonumber
\\
&+&\frac{x a}{c}
F^{2:1;1;1}_{2:0;0;1}
\left( \begin{tabular}{c}
[a+2:2,1,2]: [1:1];[b:1];[1,1]
\\
\, [c+1:1,1,1;  2:1,0,1]: [-];[-];[a+2:2]
\end{tabular}\, x,y,x \right)
\, .
\end{eqnarray}

\subsection{Lower parameter derivative}
In complete analogy to Sec.~\ref{multQnSect} the derivative of a hypergeometric function
in a lower parameter with multiple summation indices $q_\lambda n_\lambda$ reads
\begin{eqnarray}
\frac{\mathrm{d} F(b)}{\mathrm{d} b}&=&-\sum_{k,n_1,\dots,n_\phi=0}^\infty \sum_{\xi=1}^\phi \sum_{\gamma=0}^{q_\xi-1}x_\xi B(n_1,\dots,n_\xi+k+1,\dots,n_\phi)\frac{(1)_k (1)_{n_\xi}}{(2)_{n_\xi+k}}
\prod_{r=1}^\phi\frac{x_r^{n_r} x_\xi^{k}}{n_r! k!}
\nonumber
\\
&\times&
\frac{\Gamma(b)}{\Gamma(b+q_\xi)(b+\gamma)}
\prod_{r=1}^{\xi-1}\frac{(b)_{\sum_{\lambda=1}^{r-1} q_\lambda n_\lambda}  }{ (b)_{\sum_{\lambda=1}^{r} q_\lambda n_\lambda} }
\prod_{r=\xi+1}^{\phi}\frac{(b+q_\xi)_{\sum_{\lambda=1}^{r-1} q_\lambda n_\lambda+q_\xi k}  }{ (b+q_\xi)_{\sum_{\lambda=1}^{r} q_\lambda n_\lambda+q_\xi k} }
\nonumber\\
&\times&
\frac{(b)_{\sum_{\lambda=1}^{\xi-1}q_\lambda n_\lambda} }{(b+q_\xi)_{\sum_{\lambda=1}^{\xi-1}q_\lambda n_\lambda+q_\xi k} }
\frac{(b+\gamma)_{\sum_{\lambda=1}^{\xi-1}q_\lambda n_\lambda+q_\xi k} }{(b+\gamma+1)_{\sum_{\lambda=1}^{\xi-1}q_\lambda n_\lambda+q_\xi k} }
\, .
\label{multSumEqLowMult}
\end{eqnarray}
From eqs.~(\ref{multSumEqUp}) and (\ref{multSumEqLowMult}) one can then deduce
that any derivative of a generalized Lauricella series in one of its parameters
with summation indices of the type $q_\lambda n_\lambda$, $q_\lambda \in \mathbb{N}$
are expressible in terms of function in the same class.
In particular, we note that the derivative of a generalized Lauricella
function in $m$ variables can be expressed
as a finite sum of generalized Lauricella functions in $m+1$ variables.

\section{Derivative in parameter with a negative summation index}
\label{SectNeg}
\subsection{Upper parameter derivative}
If one of the parameters in a Pochhammer symbol is connected with multiple
summation indices of which one is negative, such as in the case $(a)_{n_1-n_2}$,
the derivative with respect to this parameter requires additional care.
In a first step, we can factorize
\begin{gather}
(a)_{n_1-n_2}=(a+n_1)_{-n_2}(a)_{n_1}
\, .
\label{splitting3}
\end{gather}
Here, the negative summation index $(-n_2)$ appears and eq.~(\ref{pochhammerPos}) needs to be replaced by
\begin{equation}
\Psi(z-n)-\Psi(z)=-\sum_{k=0}^{n-1}\frac{1}{z-k-1}
\, .
\label{pochhammerNeg}
\end{equation}
Then, the derivative of a Pochhammer symbol can be written as
\begin{gather}
\frac{\mathrm{d} (a)_{-n}}{\mathrm{d} a}=-(a)_{-n}\sum_{k=0}^{n-1}\frac{1}{a-1}\frac{(a-1)_{-k}}{(a)_{-k}}
\, .
\end{gather}
As an example, we consider the hypergeometric function with a
summation index of type $(a)_{m-n}$
\begin{gather}
F(a)=\sum_{m,n=0}^\infty B(n,m) (a)_{m-n} \frac{x^n y^m}{n! m!}
\, ,
\end{gather}
and by using the splitting formula in eq.~(\ref{splitting3}) together with eq.~(\ref{pochhammerNeg})
for the differentiation of a Pochhammer symbol with a negative index
and the interchange of the order of summations in eq.~(\ref{resum}) we obtain
\begin{eqnarray}
\frac{\mathrm{d} F(a)}{\mathrm{d} a}&=&y\sum_{k,n,m=0}^\infty B(n,m+k+1)\frac{(1)_k (1)_m}{(2)_{m+k}} \frac{(a)_k  (a+1)_{m+k-n}}{(a+1)_{k}}\frac{x^n y^m y^k}{n! m! k!}
\nonumber \\
&-&x\sum_{k,n,m=0}^\infty B(n+k+1,m)\frac{(1)_k (1)_n}{(2)_{n+k}} \frac{(a-1)_{m-n-k}  (a-1)_{m-k}}{(a)_{m-k} (a-1)^2}\frac{x^n y^m x^k}{n! m! k!}
\, .
\end{eqnarray}
As another example for illustration we calculate the derivative of the Horn
hypergeometric function in two variables  $H_1(a,b,c,d,x,y)$ with respect to its upper parameter $a$ with summation indices $m-n$:
\begin{eqnarray}
\frac{\mathrm{d} H_1(a)}{\mathrm{d} a}&=&-\frac{ybc }{(a-1)^2}\sum_{m,n,k=0}^\infty \frac{(1)_k (1)_{n}}{(2)_{n+k}}\frac{x^m y^n y^k}{m!n!k!}
\frac{(a-1)_{m-k}(a-1)_{m-n-k}(b+1)_{m+n+k}(c+1)_{n+k}}{(a)_{m-k} (d)_{m}}
\nonumber
\\
&+&\frac{x b }{d}\sum_{m,n,k=0}^\infty \frac{(1)_k (1)_{m}}{(2)_{m+k}}\frac{x^m y^n x^k}{m!n!k!}
\frac{(a)_{k}(1+a)_{m-n+k}(b+1)_{m+n+k} (c)_n}{(1+a)_{k} (d+1)_{m+k}}
\, ,
\end{eqnarray}
which can be written as a sum of two generalized Lauricella series,
\begin{eqnarray}
\frac{\mathrm{d} H_1(a)}{\mathrm{d} a}
&=&-\frac{ybc}{(a-1)^2}
\nonumber
\\
&\times&F^{4:0;1;1}_{2:1;0;0}
\left( \begin{tabular}{c}
[a-1:1,-1,-1; b+1:1,1,1;c+1:0,1,1;a-1:1,0,-1]: [-];[1:1];[1:1]
\\
\, [a:1,0,-1; 2:0,1,1;]: [d,1];[-];[-]
\end{tabular}\, x,y,y \right)
\nonumber
\\
&+&\frac{x b}{d}
F^{2:1;1;2}_{2:0;0;1}
\left( \begin{tabular}{c}
[a+1:1,-1,1;b+1:1,1,1]: [1:1];[c:1];[1:1;a:1]
\\
\, [d+1:1,0,1;  2:1,0,1]: [-]; [-];[a+1:1]
\end{tabular}\, x,y,x \right)
\, .
\end{eqnarray}

\subsection{Lower parameter derivative}
Following the same procedure as above for the derivative acting on lower parameter
of a hypergeometric function, we arrive at the explicit relation for the case
with a summation index of type $1/(b)_{m-n}$,
\begin{gather}
F(b)=\sum_{m,n=0}^\infty B(n,m) \frac{1}{(b)_{m-n}} \frac{x^n y^m}{n! m!}
\, .
\end{gather}
By using the same eq.~(\ref{splitting3}) and the derivative of a Pochhammer symbol
\begin{gather}
\frac{\mathrm{d} }{\mathrm{d} b} \frac{1}{(b+m)_{-n}}=\frac{1}{(b+m)_{-n}}\sum_{k=0}^{n-1}\frac{1}{b-1}\frac{(b-1)_{m-k}}{(b)_{m-k}}
\, ,
\end{gather}
and, after exchanging the order of summation, we obtain
\begin{eqnarray}
\frac{\mathrm{d} F(b)}{\mathrm{d} b}&=&-y\sum_{k,n,m=0}^\infty B(n,m+k+1)\frac{(1)_k (1)_m}{(2)_{m+k}} \frac{1}{b^2}\frac{(b)_k  }{(b+1)_{m+k-n} (b+1)_{k}}\frac{x^n y^m y^k}{n! m! k!}
\nonumber \\
&+&x\sum_{k,n,m=0}^\infty B(n+k+1,m)\frac{(1)_k (1)_n}{(2)_{n+k}} \frac{(b-1)_{m-k}  }{(b-1)_{m-n-k} (b)_{m-k} }\frac{x^n y^m x^k}{n! m! k!}
\, .
\end{eqnarray}

\section{Derivative in parameter with summation index $qn$, $-q\in \mathbb{N}$}
\label{SectQNeg}
\subsection{Upper parameter derivative}
The final step for obtaining the full set of relations for the derivative of a
hypergeometric series in one of its parameters with any set of summation
indices consists of elaborating the case of negative summation indices $qn$,
where $-q\in \mathbb{N}$,
\begin{gather}
F(a)=\sum_{n=0}^\infty B(n) (a)_{qn} \frac{x^n}{n!}
\, .
\label{qnUpdef}
\end{gather}
The derivative of the Pochhammer symbol in this case reads,
\begin{gather}
\frac{\mathrm{d} (a)_{qn}}{\mathrm{d} a}=-(a)_{qn}\sum_{k=0}^{-qn-1}\frac{1}{a-1}\frac{(a-1)_{-k}}{(a)_{-k}}
\, ,
\end{gather}
and we note that upper limit of the summation is indeed positive due to $q<0$.
Upon adapting the interchange of the order of summation in eq.~(\ref{resumqn})
to the case of negative $q$, we obtain for the derivative of a function with
summation indices of the type $qn$ with integer $q<0$,
\begin{eqnarray}
\frac{\mathrm{d} F(a)}{\mathrm{d} a}&=&-x\sum_{\lambda=0}^{-q-1}\frac{1}{a-1-\lambda}\frac{\Gamma(a+q)}{\Gamma(a)}\sum_{k,n=0}^\infty B(n+k+1)\frac{(1)_k (1)_n}{(2)_{n+k}}
\nonumber\\
&\times&
 \frac{{(a+q)_{qn+qk}}(a-1-\lambda)_{qk}  }{(a-\lambda)_{qk}}\frac{x^n  x^k}{n! k!}
\, .
 \label{qnsummEq}
\end{eqnarray}

\subsection{Lower parameter derivative}
For the same summation index appearing in a lower parameter, i.e.,
exchanging $(a)_{qn} \to 1/(b)_{qn}$ in eq.~(\ref{qnUpdef}), one obtains for the
derivative
\begin{eqnarray}
\frac{\mathrm{d} F(b)}{\mathrm{d} b}&=&x\sum_{\lambda=0}^{-q-1}\frac{1}{b-\lambda-1}\frac{\Gamma(b)}{\Gamma(b+q)}\sum_{k,n=0}^\infty B(n+k+1)\frac{(1)_k (1)_n}{(2)_{n+k}}
\nonumber\\
&\times&
 \frac{(b-\lambda-1)_{qk}  }{{(b+q)_{qn+qk}}(b-\lambda)_{qk}}\frac{x^n x^k}{n! k!}
\, .
\end{eqnarray}

\section{Convergence of the series for derivatives of hypergeometric functions}
\label{SectConv}
In establishing the region of convergence of the series for the derivatives of
hypergeometric functions we follow the same rule as in~\cite{ShrivKarls}
(see, in particular, p.56 in~\cite{ShrivKarls}),
namely, we exclude any exceptional values of the parameters,
i.e., those values, for which the series terminates, becomes meaningless or
reduces to a finite sum of hypergeometric series of lower dimension.

The region of convergence for the series which have been obtained for derivatives of hypergeometric functions
we can utilize the parameter cancellation theorem (see \cite{ShrivKarls}, chapter 4, p. 108).
This states that the region of convergence for a hypergeometric series is independent of the parameters,
provided exceptional values of parameter are being excluded.
For example, the series
\begin{gather}
\sum_{n=0}^\infty B(m,n) \frac{(a)_{qn}}{(b)_{qn}} \frac{x^n}{n!}
\, ,
\end{gather}
and
\begin{gather}
\sum_{n=0}^\infty B(m,n)\frac{x^n}{n!}
\, ,
\end{gather}
have the same region of convergence.
By using this theorem we can exclude from the hypergeometric series
all Pochhammer symbols with different parameters but with the same summation index.

As an example, we consider here explicitly the region of convergence for the
derivatives in Sec.~\ref{qnOneSect}.
The other cases can be dealt with in a the similar way.
It is easy to see that the region of convergence for the series in
eq.~(\ref{qnsummEqOne}) after application of the convergence theorem is
equivalent to the one of the series
\begin{gather}
\sum_{n,k=0}^\infty B(n+k+1) (a)_{qn+qk} \frac{x^n x^k}{(n+k)!}
\, .
\label{regConv1}
\end{gather}
Then, applying the summation formula in eq.~(\ref{resumqn}) one finds that
the region of convergence of the series in eq.~(\ref{regConv1}) is the same as
the one of the expression
\begin{gather}
\sum_{n=0}^\infty\sum_{k=0}^{qn+q-1}  B(n) (a)_{qn} \frac{x^n}{(n)!}
=\sum_{n=0}^\infty   (q n+q-1)  B(n) (a)_{qn} \frac{x^n}{(n)!}
\, ,
\end{gather}
and due to convergence theorem the region of convergence of this last series coincides with the one of
the original function before differentiation.

In this way we prove the theorem that the equations for the derivatives of
hypergeometric series have the same region of convergence as the initial
hypergeometric functions.

The region of convergence for hypergeometric series in two, three and more
variables can be found by using the Horn's theorem on convergence~\cite{ShrivKarls}
(see, in particular, p.56 in~\cite{ShrivKarls}).
This theorem implies an absolute region of convergence of those hypergeometric series,
so that the use of the summation formulae in eqs.~(\ref{resum}), (\ref{resum2n}) and
(\ref{resumqn}) in our calculations is mathematically rigorous.

\section{Derivative in parameter for the general case of summation indices}
\label{mathRes}
The combination of eqs.~(\ref{multSumEqUp}) and (\ref{qnsummEq})
finally leads to an expression for the derivative of a hypergeometric function
in one of its upper parameters related to a summation index with any integer
coefficient. The relevant equation reads
\begin{gather}
F(a)=\sum_{n_1,\dots,n_\phi=0}^\infty B(n_1,\dots,n_\phi) (a)_{\sum_{r=1}^\phi q_\lambda n_\lambda}\prod_{r=1}^\phi \frac{x^{n_r} }{n_r!}
\, , \quad q \in \mathbb{Z} \, ,
\end{gather}
\begin{eqnarray}
\frac{\mathrm{d} F(a)}{\mathrm{d} a}&=&\sum_{k,n_1,\dots,n_\phi=0}^\infty \sum_{\xi=1}^\phi \sum_{\gamma=0}^{|q_\xi|-1} x_\xi B(n_1,\dots,n_\xi+k+1,\dots,n_\phi)\frac{(1)_k (1)_{n_\xi}}{(2)_{n_\xi+k}}
\prod_{r=1}^\phi\frac{x_r^{n_r} x_\xi^{k}}{n_r! k!}
\nonumber
\\
&\times&
\frac{\Gamma(a+q_\xi)}{\Gamma(a)}
\prod_{r=1}^{\xi-1}\frac{(a)_{\sum_{\lambda=1}^{r} q_\lambda n_\lambda}  }{ (a)_{\sum_{\lambda=1}^{r-1} q_\lambda n_\lambda} }
\prod_{r=\xi+1}^{\phi}\frac{(a+q_\xi)_{\sum_{\lambda=1}^{r} q_\lambda n_\lambda+q_\xi k}  }{ (a+q_\xi)_{\sum_{\lambda=1}^{r-1} q_\lambda n_\lambda+q_\xi k} }
\nonumber\\
&\times&
\frac{(a+q_\xi)_{\sum_{\lambda=1}^{\xi}q_\lambda n_\lambda+q_\xi k} }{(a)_{\sum_{\lambda=1}^{\xi-1}q_\lambda n_\lambda} }\beta
\, ,
\nonumber
\\
\beta&=&\frac{1}{a+\gamma}\frac{(a+\gamma)_{\sum_{\lambda=1}^{\xi-1}q_\lambda n_\lambda+q_\xi k} }{(a+\gamma+1)_{\sum_{\lambda=1}^{\xi-1}q_\lambda n_\lambda+q_\xi k} }
\, ,
\quad \quad q_\xi>0\, ,
\nonumber
\\
\beta&=&-\frac{1}{a-\gamma-1}\frac{(a-\gamma-1)_{\sum_{\lambda=1}^{\xi-1}q_\lambda n_\lambda+q_\xi k} }{(a-\gamma)_{\sum_{\lambda=1}^{\xi-1}q_\lambda n_\lambda+q_\xi k} }
\, ,
\quad \quad q_\xi<0\, .
\label{multSumEqUpGen}
\end{eqnarray}
A similar equation holds for the derivative in a lower parameter,
\begin{gather}
F(b)=\sum_{n_1,\dots,n_\phi=0}^\infty B(n_1,\dots,n_\phi) \frac{1}{(b)_{\sum_{r=1}^\phi q_\lambda n_\lambda}}\prod_{r=1}^\phi \frac{x^{n_r} }{n_r!}
\, , \quad q\in \mathbb{Z}\, ,
\end{gather}
\begin{eqnarray}
\frac{\mathrm{d} F(b)}{\mathrm{d} b}&=&\sum_{k,n_1,\dots,n_\phi=0}^\infty \sum_{\xi=1}^\phi \sum_{\gamma=0}^{|q_\xi|-1} x_\xi B(n_1,\dots,n_\xi+k+1,\dots,n_\phi)\frac{(1)_k (1)_{n_\xi}}{(2)_{n_\xi+k}}
\prod_{r=1}^\phi\frac{x_r^{n_r} x_\xi^{k}}{n_r! k!}
\nonumber
\\
&\times&
\frac{\Gamma(b)}{\Gamma(b+q_\xi)}
\prod_{r=1}^{\xi-1}\frac{(b)_{\sum_{\lambda=1}^{r-1} q_\lambda n_\lambda}  }{ (b)_{\sum_{\lambda=1}^{r} q_\lambda n_\lambda} }
\prod_{r=\xi+1}^{\phi}\frac{(b+q_\xi)_{\sum_{\lambda=1}^{r-1} q_\lambda n_\lambda+q_\xi k}  }{ (b+q_\xi)_{\sum_{\lambda=1}^{r} q_\lambda n_\lambda+q_\xi k} }
\nonumber\\
&\times&
\frac{(b)_{\sum_{\lambda=1}^{\xi-1}q_\lambda n_\lambda} }{(b+q_\xi)_{\sum_{\lambda=1}^{\xi-1}q_\lambda n_\lambda+q_\xi k} }\beta
\, ,
\nonumber
\\
\beta&=&-\frac{1}{b+\gamma}\frac{(b+\gamma)_{\sum_{\lambda=1}^{\xi-1}q_\lambda n_\lambda+q_\xi k} }{(b+\gamma+1)_{\sum_{\lambda=1}^{\xi-1}q_\lambda n_\lambda+q_\xi k} }
\, ,
\quad \quad q_\xi>0
\, ,
\nonumber
\\
\beta&=&\frac{1}{b-\gamma-1}\frac{(b-\gamma-1)_{\sum_{\lambda=1}^{\xi-1}q_\lambda n_\lambda+q_\xi k} }{(b-\gamma)_{\sum_{\lambda=1}^{\xi-1}q_\lambda n_\lambda+q_\xi k} }
\, ,
\quad \quad q_\xi<0
\, .
\label{multSumEqLowGen}
\end{eqnarray}

As an example we present here the derivative of the function $G_3(a,b,x,y)$ in
its upper parameter $a$ with summation indices $2n-m$:
\begin{eqnarray}
\frac{\mathrm{d} G_3(a)}{\mathrm{d} a}&=&\frac{y(a+1) }{(b-1)}\sum_{m,n,k=0}^\infty \frac{(1)_k (1)_{n}}{(2)_{n+k}}\frac{x^m y^n y^k}{m!n!k!}
\frac{(a)_{-m+2k}(a+2)_{2n-m+2k}(b-1)_{2m-n-k}}{(a+1)_{-m+2k} }
\nonumber
\\
&+&\frac{ya }{b-1}\sum_{m,n,k=0}^\infty \frac{(1)_k (1)_{n}}{(2)_{n+k}}\frac{x^m y^n y^k}{m!n!k!}
\frac{(a+1)_{-m+2k}(a+2)_{2n-m+2k}(b-1)_{2m-n-k}}{(a+2)_{-m+2k} }
\nonumber
\\
&-&\frac{x b(b+1) }{(a-1)^2}\sum_{m,n,k=0}^\infty \frac{(1)_k (1)_{m}}{(2)_{m+k}}\frac{x^m y^n x^k}{m!n!k!}
\frac{(a-1)_{-k}(a-1)_{2n-m-k}(b+2)_{2m-n+2k}}{(a)_{-k} }
\, ,
\nonumber
\end{eqnarray}
which can be written as a sum of three generalized Lauricella series,
\begin{eqnarray}
\frac{\mathrm{d} G_3(a)}{\mathrm{d} a}
&=&\frac{y(a+1)}{b-1}
\nonumber
\\
&\times&F^{3:0;1;1}_{2:0;0;0}
\left( \begin{tabular}{c}
[a+2:-1,2,2; b-1:2,-1,-1;a:-1,0,2]:[-]; [1:1];[1:1]
\\
\, [a+1:-1,0,2; 2:0,1,1;]: [-];[-];[-]
\end{tabular}\, x,y,y \right)
\nonumber
\\
&+&\frac{y a}{b-1}
F^{3:0;1;1}_{2:0;0;0}
\left( \begin{tabular}{c}
[a+2:-1,2,2; b-1:2,-1,-1;a+1:-1,0,2]:[-]; [1:1];[1:1]
\\
\, [a+2:-1,0,2; 2:0,1,1;]: [-];[-];[-]
\end{tabular}\, x,y,y \right)
\nonumber
\\
&-&\frac{x b(b+1)}{(a-1)^2}
F^{2:1;0;1}_{1:0;0;1}
\left( \begin{tabular}{c}
[a-1:-1,2,-1;b+2:2,-1,2]: [1:1];[-];[1:1;a-1:-1]
\\
\, [2:1,0,1]: [-]; [-];[a:-1]
\end{tabular}\, x,y,x \right)
\, .
\nonumber
\end{eqnarray}

\section{Conclusion}
\label{conclusion}
\subsection{Derivative of Horn-type functions in a parameter}
\label{HornRes}
With the results of the present article
the derivatives of the following Horn-type hypergeometric functions in multiple variables $x_n$
in one of their parameters $a_j$ or $b_i$
\begin{gather}
\sum_{m_1,\dots,m_l}^\infty\prod_{i,j}\frac{(a_j)_{\sum_k^l q_k m_k}}{(b_i)_{\sum_k^l q_k m_k}}\prod_{n=1}^l\frac{x_n^{m_n}}{m_n!}
\, , \quad q_k\in \mathbb{Z}\, ,
\end{gather}
can be expressed with the help of eqs.~(\ref{multSumEqUpGen}) and (\ref{multSumEqLowGen}),
or, alternatively, as finite sums of Horn-type hypergeometric functions, where
the $n$-th derivative of a function in $m$ variables is expressed as series with $n+m$ variables.
The region of convergence for those derivatives is the same as for initial function.
Specifically, for the 34 distinct confluent and non-confluent Horn-type
hypergeometric functions in two variables the $n$-th derivatives are expressed as
Horn-type hypergeometric functions in $n+2$ variables.

Derivatives of the generalized Lauricella series, i.e., of Horn-type
hypergeometric series with summation coefficients $q_k\in \mathbb{N}$, in one
of their (upper or lower) parameters can be expressed as a finite sum
of the generalized Lauricella series, as it was explained in Sec.~\ref{multLaurich}.
In particular, the derivatives of the generalized Lauricella series in one of their parameters
with summation coefficient $q_k\in {0,1}$ can be written as a finite sum of generalized Lauricella
series with summation coefficient in the same alphabet $q_k\in {0,1}$ as detailed in Sec.~\ref{2inexsum}.
Finally, all $n$-th derivatives of generalized Appell hypergeometric
functions, generalized Kamp{\'e} de F{\'e}riet functions, generalized
hypergeometric functions in one variable are expressible at the terms of
generalized Lauricella series with summation coefficients $q_k\in {0,1}$.

\subsection{Applications in high-energy physics}
\label{application}
The main motivation of the present research are calculations in quantum field
theory, i.e., of Feynman diagrams and the series expansions which stem from it.
By applying the Mellin-Barnes method, Feynman diagrams can be
written in the form of eq.~(\ref{Horn}) as Horn-type multi-variable
hypergeometric functions (see e.g.,~\cite{Bytev:2009kb}).
For example, particular types of one-loop diagrams are expressible in terms of
generalized hypergeometric functions or Appell series~\cite{Dav1}--\!\cite{Dav5}.

For the evaluation of the finite part of dimensionally regularized Feynman
integrals in $D=4-2\varepsilon$ dimensional space-time one has to construct
the expansion in $\varepsilon$.
This has motivated us to seek a general method to obtain the all-order
$\varepsilon$ expansion of Horn-type hypergeometric function.
There exist a lot of analytical and numerical methods based on different
algorithms which are applicable for dealing with $\varepsilon$ expansion of
Feynman integrals~\cite{epsExp1}--\!\cite{epsExp7}
and are implemented in computer packages \cite{Moch:2001zr},
\cite{epsExpPack1}--\!\cite{epsExpPack6}.

Feynman integrals written in the form of Horn-type hypergeometric function
as in eq.~(\ref{Horn}) depend on space-time parameter $\varepsilon$ only
through some parameter in the Pochhammer symbols, namely  $B_a, D_a$,
so that the construction of the $\varepsilon$ expansion of a given Feynman integral
is equivalent to taking the derivative of Horn-type hypergeometric functions
in its parameters.

From Sec.~\ref{HornRes} we can conclude that the $\varepsilon$ expansion
of Feynman integrals at any order are expressible in terms of Horn-type
hypergeometric functions.
In detail, the $n$-th term of the $\varepsilon$ series can be expressed as
Horn-type hypergeometric function in $n+m$ variables,
where $m$ is the number of summations in Horn-type representation of the Feynman integral.
The region of convergence of any of those parameter derivatives, i.e., the
coefficient in the $\varepsilon$ expansion, and the initial Feynman integral are the same.
The explicit formula for $n$-th term of $\varepsilon$ expansion can be obtained
with the help of eqs.~(\ref{multSumEqUpGen}) and (\ref{multSumEqLowGen}).

In the series of the papers \cite{Bytev:2009kb}, \cite{my1}--\!\cite{my4} the method of
differential reduction of hypergeometric function has been worked out.
In particular, the so-called step-up and step-down differential operators
have been introduced, which shift the parameters of hypergeometric functions by unity.
By applying such differential operators to a hypergeometric function, the value of any
parameter can be shifted by an arbitrary integer. The construction of
differential operators allows to define a set of exceptional parameters for
hypergeometric function and then find the condition of reducibility of the
monodromy group of the corresponding hypergeometric functions.

By expressing the $\varepsilon$ expansion of Feynman diagram in terms of
Horn-type hypergeometric functions and applying the above mentioned method of
differential reduction one can reduce the corresponding integrals to some subset
of basic hypergeometric functions and express them as a series with the least
number of infinite summations.

\section*{Acknowledgments}

We are grateful to M.Yu.~Kalmykov for fruitful discussions, useful remarks and valuable contributions to this paper.
The work of V.V.B. was supported in part by the Heisenberg--Landau Program.
This work was supported in part by the German Federal Ministry for Education
and Research BMBF through Grant No.\ 05H15GUCC1 and by the German Research
Foundation DFG through the Collaborative Research Center SFB~676
{\it Particles, Strings and the Early Universe -- The Structure of Matter and Space-Time}.

\appendix

\renewcommand{\theequation}{\ref{hypGenDef}.\arabic{equation}}
\setcounter{equation}{0}
\section{Definition of hypergeometric series}
\label{hypGenDef}
Here we give the definitions of some hypergeometric series, whose derivatives
have been considered in the article.

\subsection{Appell hypergeometric functions of two variables}
\label{AppellDef}
The Appell~\cite{appell1,appell2} hypergeometric functions
$F_1$, $F_2$, $F_3$ and $F_4$ are defined in an expansion around $x=y=0$ as
\begin{eqnarray}
F_1(a, b_1, b_2, c; x,y)&=&
\sum_{m=0}^\infty
\sum_{n=0}^\infty
\frac{(a)_{m+n} (b_1)_m (b_2)_n}{(c)_{m+n}}
\frac{x^m}{m!}
\frac{y^n}{n!}
\, ,
\nonumber \\
F_2(a, b_1, b_2, c_1, c_2 ; x,y)
&=&
\sum_{m=0}^\infty
\sum_{n=0}^\infty
\frac{(a)_{m+n} (b_1)_m (b_2)_n}{(c_1)_m (c_2)_n}
\frac{x^m}{m!}
\frac{y^n}{n!}
\, ,
\nonumber \\
F_3(a_1, a_2, b_1, b_2, c; x,y)
&=&
\sum_{m=0}^\infty
\sum_{n=0}^\infty
\frac{(a_1)_{m} (a_2)_{n} (b_1)_m (b_2)_n}{(c)_{m+n} }
\frac{x^m}{m!}
\frac{y^n}{n!}
\, ,
\nonumber \\
F_4(a, b, c_1, c_2; x,y)
&=&
\sum_{m=0}^\infty
\sum_{n=0}^\infty
\frac{(a)_{m+n} (b)_{m+n} }{(c_1)_m (c_2)_n}
\frac{x^m}{m!}
\frac{y^n}{n!}
\, .
\label{def:f1-f4}
\end{eqnarray}

\subsection{Multi-variable extension of Kamp{\'e} de F{\'e}riet series}
\label{CamdeFerDef}
The extension of Kamp{\'e} de F{\'e}riet functions in two variables to
multi-variable case follows~\cite{ShrivKarls}
\begin{eqnarray}
F^{p:q_1;\dots q_n}_{l:m_1;\dots m_n} \left( \begin{array}{c}
x_1 \\
\vdots\\
x_n
\end{array} \right)
&=&F^{p:q_1;\dots q_n}_{l:m_1;\dots m_n}
\left[ \begin{array}{c}
(a)_p: (b^{(1)}_{q_1});\dots; (b^{(n)}_{q_n}) \\
(\alpha)_l: (\beta^{(1)}_{m_1});\dots; (\beta^{(n)}_{m_n})
\end{array}\, x_1,\dots x_n \right]
\nonumber\\
&=&\sum_{s_1,\dots,s_n=0}^{\infty}\Lambda(s_1,\dots,s_n)\frac{x_1^{s_1}}{s_1!}\frac{x_n^{s_n}}{s_n!}
\, ,
\end{eqnarray}
where
\begin{eqnarray}
\Lambda(s_1,\dots,s_n)
=\frac{\prod_{j=1}^p (a_j)_{s_1+\dots s_n}  \prod_{j=1}^{q_1} (b^{(1)}_j)_{s_1}\dots \prod_{j=1}^{q_n} (b^{(n)}_j)_{s_n} }
      {\prod_{j=1}^l (\alpha_j)_{s_1+\dots s_n}  \prod_{j=1}^{m_1} (\beta^{(1)}_j)_{s_1}\dots \prod_{j=1}^{m_n} (\beta^{(n)}_j)_{s_n} }
\, .
\end{eqnarray}

\subsection{Generalized Lauricella series}
\label{LaurGenDef}
Series of this type have been introduced in~\cite{SrivDao}.
Special cases of these functions are reduced to the multi-variable extension of Kamp{\'e} de F{\'e}riet series.
The latter includes confluent and non-confluent Lauricella functions.
\begin{eqnarray}
F^{A:B^{(1)};\dots B^{(n)}}_{C:D^{(1)};\dots D^{(n)}} \left( \begin{array}{c}
x_1 \\
\vdots\\
x_n
\end{array} \right)
&=&F^{A:B^{(1)};\dots B^{(n)}}_{C:D^{(1)};\dots D^{(n)}}
\left(
\begin{array}{c}
[(a):\theta^{(1)},\dots,\theta^{(n)}]
   \\
\, [(c):\psi^{(1)},\dots,\psi^{(n)}]
\end{array}
\right.
\nonumber\\
&& \quad\quad\quad\quad\quad\quad
\left.
\begin{array}{c}
 : [(b^{{1}}):\phi^{(1)}];\dots;[(b^{{n}}):\phi^{(n)} ]
   \\
\,  : [(d^{{1}}):\delta^{(1)}];\dots;[(d^{{n}}):\delta^{(n)} ]
\end{array}
\, x_1,\dots x_n \right)
\nonumber\\
&=&\sum_{s_1,\dots,s_n=0}^{\infty}\Omega(s_1,\dots,s_n)\frac{x_1^{s_1}}{s_1!}\frac{x_n^{s_n}}{s_n!}
\, ,
\end{eqnarray}
where
\begin{eqnarray}
\Omega(s_1,\dots,s_n)
=\frac{\prod_{j=1}^A (a_j)_{s_1\theta_j^{(1)}+\dots s_n\theta_j^{(n)}}  \prod_{j=1}^{B^{(1)}} (b^{(1)}_j)_{s_1\phi_j^{(1)}}\dots \prod_{j=1}^{B^{(n)}} (b^{(n)}_j)_{s_n\phi^{(n)}_j} }
      {\prod_{j=1}^C (c_j)_{s_1\psi_j^{(1)}+\dots s_n\psi_j^{(n)}}  \prod_{j=1}^{D^{(1)}} (d^{(1)}_j)_{s_1\delta_j^{(1)}}\dots \prod_{j=1}^{D^{(n)}} (d^{(n)}_j)_{s_n\delta^{(n)}_j} }
\, ,
\end{eqnarray}
and all parameters
$\theta_1^{{(1)}},\psi_1^{{(1)}},\phi_1^{{(1)}},\delta_1^{{(1)}}\dots \theta_A^{{(n)}},\psi_C^{{(n)}},\phi_{B^{(n)}}^{{(n)}},\delta_{D^{(n)}}^{{(n)}}$
are positive and real.

\subsection{Horn series in two variables}
\label{HornDef}
Here we recall the definitions for some Horn functions in two variables that
have been used in the article:
\begin{gather}
H_1(a,b,c,d,x,y)=\sum_{n,m=0}^\infty\frac{(a)_{m-n}(b)_{m+n} c_n}{(d)_{m}}\frac{x^m y^n}{m!n!}
\, ,
\end{gather}
\begin{gather}
H_3(a,b,c,x,y)=\sum_{n,m=0}^\infty\frac{(a)_{2m+n}(b)_n}{(c)_{m+n}}\frac{x^m y^n}{m!n!}
\, ,
\label{H3Def}
\end{gather}
\begin{gather}
G_3(a,b,x,y)=\sum_{n,m=0}^\infty{(a)_{2n-m}(b)_{2m-n}}{}\frac{x^m y^n}{m!n!}
\, .
\end{gather}


\end{document}